# Higher education assessment practice in the era of generative AI tools

| | | |
|---|---|---|
| Bayode Ogunleye[A] | A | School of Architecture, Technology & Engineering, University of Brighton, Brighton BN2 4GJ, United Kingdom |
| Kudirat Ibilola Zakariyyah[B] | B | School of Architecture, Technology & Engineering, University of Brighton, Brighton BN2 4GJ, United Kingdom |
| Oluwaseun Ajao[C] | C | Department of Computing & Mathematics, Manchester Metropolitan University, Manchester, M1 5GD, United Kingdom |
| Olakunle Olayinka[D] | D | Department of Computer Science, University of Sheffield, Sheffield S1 4DP, United Kingdom |
| Hemlata Sharma[E] | E | Department of Computing, Sheffield Hallam University, Sheffield S1 2NU, United Kingdom |

**Abstract**

The higher education (HE) sector benefits every nation's economy and society at large. However, their contributions are challenged by advanced technologies like generative artificial intelligence (GenAI) tools. In this paper, we provide a comprehensive assessment of GenAI tools towards assessment and pedagogic practice and, subsequently, discuss the potential impacts. This study experimented using three assessment instruments from data science, data analytics, and construction management disciplines. Our findings are two-fold: first, the findings revealed that GenAI tools exhibit subject knowledge, problem-solving, analytical, critical thinking, and presentation skills and thus can limit learning when used unethically. Secondly, the design of the assessment of certain disciplines revealed the limitations of the GenAI tools. Based on our findings, we made recommendations on how AI tools can be utilised for teaching and learning in HE.

**Introduction**

The higher education (HE) sector contributes to every nation's economy and benefits society in various ways. For example, HE contributes to political stability, greater social mobility, improved social capital, crime reduction, greater social cohesion, innovation, trust, and tolerance (Brennan et al., 2013). Over the years, HE has been known to prepare individuals for the future by equipping learners with the required skills for employment. HE provides a pathway to specific careers such as law, pharmacy, and medicine (Harvey, 2000). Unfortunately, this sector is now seen to face uncertainty due to the increasing development of advanced technologies such as GenAI tools. There has been an increasing fear about the use of AI and its effect on education (Daun & Brings, 2023). The rise of GenAI has produced innovative systems such as ChatGPT (Brown et al., 2020) and Gemini (formerly known as Bard), which have taken the world by storm. These tools were a result of innovative solutions developed using large language models (LLMs) such as the generative pre-trained transformers (GPT) series developed by OpenAI, generalised autoregressive pre-training for language understanding (XLNet) developed by Google, Salesforce's conditional transformer language models (CTRL), Google's Pathways language model (PaLM), and Meta's large language model meta-AI (LLaMa). The LLMs have been used for various natural language processing tasks such as questioning and answering (Pochiraju et al., 2023), sentiment analysis (Habbat et al., 2022), topic modelling (Ogunleye et al., 2023), cyberbullying detection (Ogunleye & Dharmaraj, 2023), and fake news detection (Caramancion, 2023). As detailed in Table 1 below, the development of LLMs is mainly dominated by a few large organisations, including Google, Meta, and Microsoft/OpenAI. This is due to the very large amount of data (parameters) used to pre-train these models and the significant computational resources required to build them. However, a growing number of these models are now existing as fine-tuned applications in open-source platforms such as Hugging Face and Stable Diffusion.

Table 1. List of popular large language models (LLMs).

| Large Language Model | No Parameters Trained on | Released By | Publication |
|---|---|---|---|
| LLaMA | 65 billion | Meta | Touvron et al. (2023) |
| LaMDA | 137 billion | Google | Thoppilan et al. (2022) |
| PaLM | 540 billion | Google | Chowdhery et al. (2023) |
| UL2 | 20 billion | Google | Tay et al. (2022) |
| Codex | 12 billion | Microsoft/OpenAI | Chen et al. (2021) |
| GPT-3 | 175 billion | Microsoft/OpenAI | Brown et al. (2020) |

It is worth mentioning that the parameters of recent LLMs like GPT-4 are undisclosed due to the competitive landscape and the safety implications of large-scale models (OpenAI, 2023a). In academia, Daun and Brings (2023) discussed the use of GenAI tools such as ChatGPT for teaching and learning. The authors inferred that the tools can be used for self-assessment of one's own solution, answering student queries, and generating exercises. Baidoo-Anu and Ansah (2023) added that the systems can support teaching and learning by providing personalised tutoring, language translation, interactive learning, and automated essay grading. While GenAI solutions can serve as a valuable resource, the technology can also be misused. For example, students can make use of the system for cheating (Cotton et al., 2024); thus several concerns have been raised about these AI tools (Chaudhry et al., 2023; Cotton et al., 2024; Farrokhnia et al., 2023; Halaweh, 2023; Rasul et al., 2023). Specifically, Rasul et al. (2023) and Nikolic et al. (2023) stated that academic integrity is potentially compromised as ChatGPT has proven its competence in achieving success on medical licensing and law exams as well as producing research abstracts/contents, statistical analyses, and computer programs that are not detectable. Some recent attempts at detecting AI-generated content include LLM watermarking models (Sadasivan et al., 2023; Tang et al., 2023) and ChatGPT Checker (OpenAI, 2023b). However, studies like Chaka (2023, 2024) show that AI content detectors are inconsistent and unreliable at identifying AI-generated content. It is increasingly shown that these content detection approaches are failing in their efforts, as the GenAI models that created them are also becoming more sophisticated (*The Times*, 2023). It has been found that fine-tuning these foundational AI models, such as Bidirectional Encoder Representations from Transformers (Devlin et al., 2018) and LLaMA (Touvron et al., 2023) on domain-specific data, makes them even more effective at creating human-like responses with closely aligning domain-relevant contexts such as BioBERT (Lee et al., 2020). This further necessitates the need, especially for HE stakeholders, to measure the impact and define ways to incorporate the AI system for beneficial use in academia.

The impact of GenAI tools on teaching, learning and assessment practice in HE is a hotly debated topic (Rudolph et al., 2023a, 2023b). The use of authentic assessment is widespread in UK universities, and the adoption of it has been praised over traditional forms of assessment as it appears to have reduced the negative effect of class size on student attainment (Richardson, 2015). However, assessment practices face more challenges due to the development of AI tools. Based on this background, few studies evaluated the capabilities of GenAI tools on assessment instruments such as examination questions, essays, and coursework. For example, Mahon et al. (2023) assessed the capabilities of ChatGPT on computer science A-level examination. Finnie-Ansley et al. (2022) evaluated the performance of OpenAI Codex on introductory programming (Python) exams. Furthermore, Bartoli et al. (2024) assessed the performance of ChatGPT on neurosurgical residents' written exams. In summary, few studies have investigated the capacity of ChatGPT by assessing the content (assessment solutions) generated. However, this is limited to medical education and programming context. There is a paucity of studies across STEM disciplines that assessed the performance of GenAI tools on assessment in the HE pedagogic practice. Thus, our study intends to fill this gap. Our work did not aim at making judgments on the use of AI-generated content in academia. However, we aim to assess the performance of GenAI tools in STEM-related disciplines to understand their potential impact on students' learning and development.

Most GenAI systems are still in an early stage of development, and there are limited studies on how solutions generated from these systems can impact assessment practice, such as assessment by coursework or essay. Thus, this paper contributes in several ways to teaching, learning and assessment practice in HE, and these contributions can be summarised as follows.

- This study demonstrates a methodology for evaluating the impact of GenAI tools towards assessment practices in HE. We show that AI tools present an opportunity to evaluate, critique and contextualise information. However, this may appear to be discipline-dependent.

- Our study shows the necessity of modifying curricula to accommodate the evolving skill sets demanded. Additionally, we emphasise the importance of providing proper guidance for the utilisation of GenAI. This entails integrating it into courses under proper supervision, rather than allowing unregulated student use, which could impact their learning experience detrimentally.

- Our findings demonstrate the potential impact of GenAI tools on students' learning when used for assessments unethically. Subsequently, we present assessment instruments and criteria which are helpful for future research and learning support in this era of AI-generated content.

The rest of the paper is organised as follows: Section 2 reviews the literature to provide background knowledge for this study. Section 3 centres on the methods, while Sections 4 and 5 present the results, conclusions, and recommendations.

**Related work**

The HE sector has been faced with challenges since the advent of Google's search service in 1997 (Brophy & Bawden, 2005). HE faculty members were overwhelmed with the fear of the tool taking their positions as learners could easily search for materials to learn online through these search engines. However, over the years, this has proven not to be the case. Search engines have been used for teaching, learning, and research effectively, and the educational sector has been one of the sectors that benefitted most from the technology. The recent GenAI tools are conversational agents that can generate human-like content such as texts, images, and videos. For illustration, GenAI applications such as ChatGPT are a form of question-answering models (Wu et al., 2023). In their development, they would usually require example questions, which are labelled datasets in diverse semantic and structural formats, sometimes domain-specific, for the GenAI applications to be pre-trained such that they can give the user the correct answers when prompted (asked a question). There are benchmark datasets that exist and have been used for this purpose; these include arithmetic-type questions (Cobbe et al., 2021) and common sense-type questions (Talmor et al., 2018). Table 2 below provides a summary of some of the benchmark datasets. It is worth stating that there are other labelled datasets that were used for training the recent LLMs like GPT-4. The datasets are bound to be much larger; however, information about them has not been disclosed. In academia, there are fears about GenAI's effect on HE (Daun & Brings, 2023). Kaplan-Rakowski et al. (2023) investigated teachers' perspectives on the use of GenAI for teaching and learning, using 147 diverse groups of teachers via an online survey. Their survey questions were around technology integration (in terms of awareness, learning, understanding, familiarity, adaptation, and application), participants' perceptions of GenAI implementation in education, and the use of GenAI for teaching. Their result showed that teachers have a positive perception of the use of AI tools. However, that does not translate into actions. Grassini (2023) discussed both the potentials and challenges associated with the integration of generated AI tools in academia. They identified key potentials of AI as able to assist in providing feedback and developing learning materials. Equally, key challenges, such as bias, hallucinations, academic integrity, and data privacy, were highlighted.

Table 2. List of benchmark datasets for pre-training LLMs.

| Reasoning Dataset | Category | Description | Publication |
|---|---|---|---|
| GSM8K | Arithmetic | Math Word Problems (8.5K) | Cobbe et al. (2021) |
| SVAMP | | Math Word Problems (1.0K) | Patel et al. (2021) |
| ASDiv | | Math Word Problems (1.2K) | Miao et al. (2021) |
| AQuA | | Algebraic Word Problems (100K) | Ling et al. (2017) |
| MAWPS | | Math Word Problems (2.4K) | Koncel-Kedziorski et al. (2016) |
| CSQA | Common Sense | Complex semantics (12.2K) | Talmor et al. (2018) |
| StrategyQA | | Multi-hop strategy (2.8K) | Geva et al. (2021) |
| BIG-bench (Date) BIG-bench (Sports) | | Inferring Date from context (204) Plausibility of sports sentences | Srivastava et al. (2022) |
| SayCan | | Mapping instructions to robot actions (101) | Ahn et al. (2022) |

Assessment instruments like examinations or coursework are tools which are used to evaluate and enhance student learning. Assessment strongly influences students' learning (Bloxham, 2015). The marking of students' assessments involves evaluating various aspects of students' performance. Some of the elements of assessment include accuracy and validity, the demonstration of learning, the transfer of knowledge, collaboration, and metacognition (Ashford-Rowe et al., 2014). Kim et al. (2019) opined that higher-order skills that cover complex thinking, communication, collaboration, and creativity, which are also referred to as the 4Cs, are the most significant authenticity or future skills criteria. Crawford et al. (2023) emphasised on demonstrating comprehension of a subject to solve complex problems as an assessment criterion rather than regurgitating theories in a textbook. Academics have used criterion-referenced assessment (CRA) successfully based on its reliability, validity, and transparency in assessing learning (Burton, 2006; Liao, 2022; Lok et al., 2016). Assessment criteria play a vital role in defining and assessing students' performance in various educational setups. The criteria offer clear, objective standards against which students' work is evaluated. The criteria must be mentioned clearly; this will help students understand what is required and how the assessment will be marked (Popham, 1997). The approach helps reduce variations in marks awarded to students. Despite the importance of assessment to learning, the availability of GenAI tools poses several threats to authentic assessment. A major concern is the student usage for cheating. This has brought a lot of attention to HE to implement policy and guidance on the usage of AI systems. Most importantly, to integrate GenAI into academia. Currently, there is no agreed guideline for the usage of GenAI systems in HE, and thus, it is worth assessing the capabilities of the systems across several disciplines to inform policy-making processes. Past studies assessed the performance of GenAI tools using assessment instruments like examination questions and coursework. For example, Malinka et al. (2023) tested ChatGPT's performance on programming tasks and concluded that ChatGPT has the capacity to pass the courses required for a university degree in IT security. Kolade et al. (2024) deployed ChatGPT to generate academic essays on the digital transformation of the health sectors in the global South with suggestions on improving digitally enabled healthcare delivery. Their study showed that ChatGPT 3.5 generates original high-quality content that is hard to distinguish from human-generated content. To conclude, we present a summary of existing key literature that assessed the performance of GenAI tools for pedagogy practice in Table 3. This is beneficial for academics to understand solutions that AI systems can generate in their discipline.

Table 3. Summary of key literature that assessed the capabilities of GenAI tools.

| Study | Aim | Method | Findings | Conclusion & Recommendation |
|---|---|---|---|---|
| Thibaut et al. (2024) | The paper assessed ChatGPT and Bard's abilities to pass the first part of the European Board of Hand Surgery (EBHS) diploma examination. | Prompts and answers from ChatGPT and Bard | The GenAI results were poor. The study showed there is no significant difference in the performance of the GenAI tools when answering the multiple-choice exam questions. However, they are still developing their learning capability. | Their study concluded that ChatGPT and Bard are not capable of passing the first part of the EBHS diploma exam in their current state. |
| Cuthbert & Simpson (2023) | The study assessed whether ChatGPT could pass Section 1 of the Fellowship of the Royal College of Surgeons (FRCS) examination in Trauma and Orthopaedic Surgery. | Prompts and answers from ChatGPT | ChatGPT scores were 30-35% lower than the FRCS pass rate, with 8.2% lower than the mean score achieved by human candidates of all training levels. ChatGPT had 53.3% in basic science but 0 % in trauma. | ChatGPT lacks the higher-order judgment and complex thinking required to pass the FRCS examination. In addition, it also fails to recognise its own limitations. ChatGPT's deficiencies should be publicised equally as much as its successes to ensure clinicians remain aware of its fallibility. |
| Mahon et al. (2023) | Assessed ChatGPT-4 abilities on computer science examinations (UK A-Level and Irish Leaving Certificate). | Prompts and answers from ChatGPT | The results indicated that ChatGPT is capable of achieving very high marks on both examinations. Furthermore, their results showed that the performance differences before and after the knowledge cut-off date (September 2021) are minimal. However, ChatGPT struggled with image and symbol questions. Minimal hallucination occurrences were observed. | The occurrence of hallucinations in answers and a few errors in the solutions provided in questions without images call for concern in examination designs. In the future, ChatGPT can be adopted to gather information and discuss tactics during an examination, and such an interactive process can be recorded. |
| Gupta et al. (2023) | The study tested the GPT-4 exploitation as an instrument for plastic surgery graduate medical education by evaluating its performance on the Plastic Surgery Inservice Training Examination (PSITE). | Prompts and answers from ChatGPT | GPT-4 answers were 77.3% accurate (187 out of 242 questions were answered correctly) | GPT-4 possesses excellent accuracy and reliability for plastic surgery resident education, over GPT-3.5. Academics/Practitioners should utilise ChatGPT to enhance their educational curriculum. |
| Bartoli et al. (2024) | The paper assessed how ChatGPT performs at both generating questions and answering a neurosurgical resident's written exam. | Prompts and answers from ChatGPT | Though the ChatGPT required an iterative process, it answered all its self-generated questions correctly, and there was no difference in response rate for residents between human-generated and AI-generated questions, which could have been attributed to the lack of clarity of the questions. | AI is a promising and powerful tool, but it should be used for specific medical purposes that need to be further determined. To enhance its versatility, the prompts must be carefully, precisely, and reasonably formulated. |
| Wang et al. (2023) | The authors evaluated ChatGPT's ability using three medical examination datasets. The testing involved ordering ChatGPT to act as a doctor to answer exam questions, provide patient discharge summaries, and provide diagnoses. | Prompts and answers from ChatGPT | GPT3 performed well in the China National Medical Licensing Examination in Chinese (CNMLE), its English version (ENMLE), and the China National Entrance Examination for Postgraduate-Clinical Medicine Comprehensive Ability (NEEPM) exams with scores of 56%, 76% and 62% respectively, However, it had lesser scores compared to GPT-4, that scored 84%, 86%, and 82%. Nevertheless, in other areas/criteria such as verbal fluency, open and close domain hallucinations, etc., both GPT 3 and GPT-4 took the turns, that is where one takes the lead, and the other takes the tail. In summary, GPT-4 appears more promising. | In general, ChatGPT accuracy was good. However, it is worth stating that the system still struggles with key medical information, insufficient diagnosis and false information. |
| Rudolph et al. (2023b) | The study compared the performances of ChatGPT, Bing chat and Bard using 15 questions. | Prompts and answers from ChatGPT, Bing chat and Bard | The study showed that ChatGPT-4 performed better than Bing chat and Bard. However, in general, the chatbots are not yet at the A or B level grade. In addition, the study revealed a multi-disciplinary test that is relevant for HE assessments. | Recommends the integration of AI systems to support learning. Academics should encourage the use of oral exams to assess student learning. Academics should endeavour to use authentic assessments for assessing student learning outcomes. Students should declare the use of AI systems in assessments |
| Flores-Cohaila et al. (2023) | The paper assessed ChatGPT 3.5 and GPT-4 accuracies on the Peruvian National Licensing Medical Examination (Examen Nacional de Medicina [ENAM]) and the identification of factors associated with incorrect answers provided by ChatGPT. The study made use of 180 multiple-choice questions from the ENAM 2022 data set. | Prompts and answers from ChatGPT | On the ENAM, GPT-4 achieved an accuracy of 86%, GPT-3.5 77%, and the 1025 examinees 55%, implying that GPT 3 and GPT 4 outperformed human candidates. Their results suggest incorrect answers were associated with the difficulty of questions, which may resemble human performance. | The study concluded that ChatGPT (GPT-3.5 and GPT-4) can achieve expert-level performance on the ENAM, outperforming most examinees. |
| Kunitsu (2023) | The study evaluated GPT-4's ability to answer questions from the Japanese National Examination for Pharmacists (JNEP). | Prompts and answers from ChatGPT | GPT-4 had an accuracy score of over 60%. | The authors concluded that though the bot accuracy rate is good, and they (ChatGPT-4) can support pharmacists' capabilities, the bots have limitations, especially in handling highly specialised questions, calculation questions, and questions requiring diagrams. |
| Parker et al. (2024) | The authors investigated the use of AI in undergraduate assessment with a focus on the ability of graduate teaching assistants (GTAs) to spot AI-generated assessments. They also examined the performance of ChatGPT in producing high-quality work. | ChatGPT | ChatGPT excelled the average student in all classes, with top marks in 8 of the 10 classes while the GTAs were only able to identify 50% of the AI-generated assessments solutions. | AI consistently performed well, indicating its robust capability in handling a wide range of assessment types and academic subjects. They equally concurred that there are variations in the literature on models of results when AI is used in assessment. |
| Funk et al. (2024). | The paper evaluated GenAI tools in terms of reliability using 450 medical examination questions. | ChatGPT 3.5 and 4 | ChatGPT-4 showed better accuracy than ChatGPT-3.5. | Though ChatGPT-4 performance outweighed ChatGPT-3.5, the significance of human factor in medical education and clinical decision making, cannot be overemphasised. |

## Methodology

This section provides details on the GenAI, and assessment tools employed. In this paper, we prepared three case studies of assessments from the data science, data analytics, and construction management disciplines (as shown in Appendix A – C). The selection of these disciplines was based on the authors' expertise. OpenAI's GPT-4 and Google's Bard (now Gemini) are well-performing GenAI tools among the LLMs that have transformed how we interact with machines and process massive amounts of text information (Dhanvijay et al., 2023). These models can generate text like human beings and perform a variety of tasks (Mohamadi et al., 2023), such as text completion, language translation, and content generation. As a result of their versatility, effectiveness, popularity, innovativeness, and usage in the literature, both of these tools are compelling choices for users across a range of industries and research fields. For these reasons, our study employed ChatGPT-4 and Bard as our GenAI tools to attempt the assessments and generate solutions to the tasks. These tools were accessed and used for our experiment within the entire month of September 2023. The assessment case studies in subject areas of data analytics, data science, and construction management were chosen at the master's degree level because all authors have taught at that level of study and are well-informed and equipped to assess work submitted for assessment in those disciplines. We used the standard grading scale (A – G) to provide marks for the solutions generated by the AI tools. The grade scale is selected because it is commonly used, especially in UK universities (University of Aberdeen, n.d.; Zarb et al., 2023). The grade scale can be interpreted as shown in Table 4.

Table 4. Interpretation of grades.

| Grade | Interpretation |
|---|---|
| A | Excellent |
| B | Very good |
| C | Good |
| D | Pass |
| E | Marginal Fail |
| F | Fail |
| G | No submission |

Furthermore, each assessment case study made use of criterion-referenced assessment due to the sound theoretical rationale, effectiveness, suitability, appropriateness, and applicability (Liao, 2022; Lok et al., 2016). Our assessment criteria are developed based on criteria/skill sets identified in the literature. Our literature findings suggest comprehension (Crawford et al., 2023; Rudolph et al., 2023b), analysis, and accuracy (Ashford-Rowe et al., 2014; Rudolph et al., 2023b). We added discussion and presentation criteria. The latter criteria were selected to assess critical thinking, coherence, hallucination, and bias. Incidentally, the criteria are common within the three disciplines. Table 5 presents the assessment criteria used for grading the solutions generated by ChatGPT and Bard (also, the marking rubric can be seen in Appendix D). For evaluation purposes, the assessments and marks awarded were moderated by authors and subject-area colleagues. We involved at least two academic tutors for each case study to moderate the assessment instruments and the marks awarded. This is to ensure instructions are clear and the level of difficulty suits the level of a master's degree. In addition, the approach aims to minimise bias and variance in marks awarded.

Table 5. Assessment criteria.

| Assessment Criteria | Assessment Skills Outcome | Description | Grade (Scale: A-G) |
|---|---|---|---|
| Comprehension and Application | Subject Knowledge, Problem Solving | Understanding of the problem, explanation, Interpretability, and use of literature from a wide range of disciplines. | |
| Analysis and Synthesis, Accuracy and Relevance | Innovative/Creativity, Problem Solving | Reasoned debate and analysis, and use of theory relevant to the subject. | |
| Evaluation/Discussion of Result | Critical Thinking | Critical evaluation demonstrated. Implications of the critique identified. | |
| Presentation/ Readability | Coherence, communication, organisation, independent learning | Coherent use of language, structure, presentation and referencing. | |

**Result and discussion**

The marks awarded to the solution provided by ChatGPT and Bard for each of the assessments are presented in Table 6. Marks were awarded by authors with subject knowledge and moderated by at least two other academic tutors in the field. The first assessment is a data analytics problem that involves a binary classification task in the context of heart failure prediction to make informed decisions for a patient's survival. The tasks involve data exploration, fitting machine learning (ML) algorithms with imbalanced and balanced dataset, evaluation of the ML algorithms performance using appropriate metrics and lastly, recommendation of the best performing model.

Table 6. Grades awarded.

| | Assessment Number | 1 | 1 | 2 | 2 | 3 | 3 |
|---|---|---|---|---|---|---|---|
| | GenAI Model | ChatGPT | Bard | ChatGPT | Bard | ChatGPT | Bard |
| Assessment Criteria | Comprehension and Application | B | C | C | B | B | D |
| | Analysis and Synthesis, Accuracy and Relevance | A | B | B | B | C | D |
| | Evaluation/Discussion of Result | B | C | C | A | D | E |
| | Presentation/Readability | B | B | C | B | D | D |

The solution generated by ChatGPT shows that the system possesses fine-grained interactive features, subject knowledge, problem-solving, analytical, critical thinking, and presentation skills. The report shows a coherent discussion of the given tasks. For example, the system provided the metadata, performed descriptive statistics, and performed data pre-processing, like the missing values was checked, and feature scaling was done. Surprisingly, the system was able to identify the target variable without stating it and identified the number of classes of the target variable (binary in this case). The system applied the oversampling technique to rebalance the minority class. Afterwards, all the ML algorithms were fitted appropriately. The system offered the flexibility to choose the proportion of data for training and testing purposes. Thereafter, the performance of the ML algorithms was evaluated using metrics such as accuracy, precision, recall, and F1-score. In addition, the system produced bar plots of the ML performance result and discussed the importance of the evaluation metrics in detail. Furthermore, the system provided a critical discussion of why a simple and interpretable model is preferred in the medical domain. However, it is worth stating that the system failed to use the "imblearn" library (Python module) to implement

SMOTE (synthetic minority oversampling technique) technique for rebalancing the class distribution. Also, the solution did not provide any supporting justification.

Similarly, the solution generated by Bard indicates that the system 'possesses' subject knowledge, problem-solving, and analytical skills. The report shows some discussion of the given task. For illustration, the system provided the metadata, performed descriptive statistics, and performed data pre-processing, such as the missing values and feature scaling. Also, the system was able to identify the target variable and the number of classes of the target variable. In comparison to the ChatGPT solution, Bard was able to utilise the SMOTE (synthetic minority oversampling technique) technique to rebalance the class distribution of the target variable. The system used 75% of the data for training and 25% for testing purposes. Afterwards, all ML algorithms were fitted, and the performance was evaluated using appropriate metrics such as accuracy, precision, recall, and F1-score. Surprisingly, all models produced from Bard showed better performance than results produced from ChatGPT. In summary, for Assessment 1, both GenAI tools were able to solve the problem provided. However, unlike ChatGPT, Bard provided references, links to source code and materials useful for the analysis/further reading. These findings are consistent with the study of Kolade et al. (2024), which found that ChatGPT struggles with referencing in assessment tasks. Furthermore, Bard appears to be technically better than ChatGPT. However, it is limited in its interactive ability, flexibility, critical thinking, discussions, and presentation of results.

The second assessment task involved a data science problem that required a reasonable amount of critical thinking. The task requirement was to provide multi-class (of three levels) text classification of a high-dimensional cancer dataset. Due to the size of this dataset, only the URL link was given in the task specification. This was presented in the case of a conventional ML task, where students may be required to pull data from some repository or external database to provide the solution to their work. A further requirement was the suggestion of using a transformer model such as BERT to convert the text into an embedding space before classification. Finally, there was the need to offer an open-ended recommendation for improving model performance in the context of providing a biomedical solution.

Although in terms of analysis, synthesis, accuracy, and relevance, Bard tended to have a similar score. ChatGPT gave a theoretical solution and pseudocode for solving the task and a good attempt to explain the steps involved. Bard gave a well-written comprehensive code but relatively smaller snippets of text in between. For evaluation and discussion of results, Bard gave well-detailed evaluation results of precision, recall, f-measure, and accuracy with impressive results attained. However, ChatGPT could not go further to give a realistic solution due to its limited ability to handle big data. Bard, on the other hand, discussed various ways of improving model performance and optimisation, including the use of biomedical data to further fine-tune the transformer model. ChatGPT presented the results of the tasks in a more organised fashion, having well-formatted and numbered sections. However, as the solution for the task was not completed, significant marks were lost here. Overall, in providing the solution to Assessment 2, it is observed that Bard appeared to have given a comprehensive solution in its implementation, while ChatGPT has provided more context, albeit in a much less robust technical analysis of the problem. In terms of comprehension and application, Bard tended to show greater coverage of the task requirement and more attention to detail. We are unsure if it was due to the nature of the case study being more data-intensive or the use of LLMs, which required significant computational power.

The third assessment is a construction management problem. The solution by ChatGPT is commendable. The information provided showed that the system possesses the subject knowledge, understanding and problem-solving skills and showed innovativeness and creativity in solving the problem. However, one schematic that takes care of all the assessment design criteria, as demanded by the assessment design, would have sufficed. We noted that some important information needed was not provided. For example, the purpose of the meeting for which the slide became a requirement was entirely omitted or ignored. Secondly, the context, that is, the Qukzome Community Health Center, was not considered. In addition, the title, as proposed by the GenAI, failed to take cognisance of the variables or themes mentioned in the assignment design. Consequently, dispute, a key variable in the assessment, was not reckoned in the slide suggested by ChatGPT. This implies that critical analysis, thinking, and evaluation were not taken care of. Similarly, Bard's solution did not capture information on how the one-slide PowerPoint presentation would be delivered, though its detailed explanations can be valuable. However, the absence of information or guidance on the slide presentation, with regards to the Iron Triangle and other constraints, reveals that Bard lacks complete comprehension of the task required, and so limits problem-solving efficiency. Bard failed to show either innovativeness or creativity in solving this problem. There was a complete absence of any schematic that recognises all the assessment design criteria.

The key criteria for critical thinking and evaluation, as omitted by ChatGPT, were equally omitted by Bard. Though Bard gave some interrelationships among the criteria mentioned, the information on the slide and the criteria were poorly presented. This is a clear indication that critical analysis, thinking, and evaluation of the assessment design was completely missing. Thus, it falls below that of ChatGPT, which gave an illustration of the iron triangle. In general, both solutions provided are coherent, organised, and well-communicated. However, the solutions did not emphasise the need to "avoid or reduce dispute", another key factor that buttressed the two AI tools' limitations for problem/project-specific or case-specific assessment. The assessment design did not request that references be provided, so both AI tools were graded equally on this aspect.

Our findings are two-fold. First, overall, the results from two of the assessments evidenced that GenAI tools possess interactive features (communication), subject knowledge, problem-solving, analytical, discipline-dependent critical thinking, research, and presentation skills. These are the essential skills that the assessments aim to measure (Lok et al., 2016). For the construction management-related assessment, the tools struggled with complex thinking skills. This is an indication that the system performance varies across the disciplines. This is due to the availability of learning resources. The availability of learning resources online varies across disciplines, which impacts the amount of content on which the LLMs were trained. The data science and data analytics disciplines encourage open learning approaches such as open sources, boot camps, and open access. Thus, there are many publicly available learning resources and opportunities. The LLMs were trained with more resources in these disciplines and, thus, can generate more content when asked. This is evidenced by the links provided by Bard. Several links to relevant online solutions, including source codes (for example, GitHub links), were produced by Bard. In practice, this implies that relevant and largely accurate content can be generated in the field of data science and data analytics for teaching and learning with a low level of human instructors for intervention. However, fields like construction management are not yet at that stage.

Considering the current state of teaching and learning in HE, we establish that GenAI tools can limit the development of learning and employability skills when used unethically, specifically in areas of data analytics and data science. This is because solutions provided by the systems showed a fine level of accuracy and relevance. However, the performances of the GenAI tools vary across disciplines, as evidenced by the result from the third assessment. For this class of disciplines, the more project-based the assessment design or practice is, the more difficult it becomes for the GenAI tools. Thus, there is a need for future research to consider experimenting in other disciplines to improve generalisation as our results are limited to the fields of data analytics, data science and construction management. Based on our results, we recommend that tutors urgently consider redesigning the assessment instrument considering the information students can develop from the GenAI tools. Similar to the results of Srivastava et al. (2022), our analysis suggests that when GenAI is used unethically, learning and development of critical thinking skills are hindered.

## Conclusions and recommendations

In this study, we aimed to assess the performance of GenAI tools in STEM-related disciplines to understand their potential impact on students' learning and development. We prepared three case study assessments in the data analytics, data science, and construction management disciplines using ChatGPT and Bard as our GenAI tools. Our results showed that GenAI tools possess subject knowledge, problem-solving, analytical, critical thinking, and presentation skills and, thus, can limit the development of students' learning when used unethically. However, this is discipline-dependent, as two sets of results emerged. It is worth stating that there were minimal occurrences of hallucination and bias in the solutions provided.

In practice, we recommend that the HE sector takes an urgent step in incorporating both GenAI systems for teaching and learning and academic AI content detector features into the plagiarism-detecting system. This is because AI content detectors (as a stand-alone system) appear to still be in the process of refinement, lacking the capability to differentiate between AI-generated content and human-written text consistently and convincingly (Chaka, 2023, 2024). Thus, our recommendation agrees with Chaka (2024), who proposed the use of both AI content detectors and plagiarism detection tools together with human reviewers. Furthermore, there is a need to re-design assessments. Academic tutors need to get familiar with GenAI systems and thus ensure authentic assessments are prepared to limit students' use of generating solutions from the AI systems. This can be achieved by strategically contextualising assessment. In addition, we encourage the use of presentation as a tool to evidence student learning outcomes. Interestingly, this can be achieved via different formats, such as in-person, virtual, or video recording. Alternatively, we recommend that tutors include AI-generated solutions (including the variants) in the assessment brief, and the assessment task can be in the form of a reflective learning approach. In this case, students can produce a report that critically reflects on the GenAI outputs (solutions) to the assessment task and proffer solutions for optimality.

Similarly, for student engagement, we recommend the use of the GenAI systems as an interactive tool during teaching and learning sessions to stimulate the student learning environment. This can be achieved in various ways. For example, students can engage in a comparative activity (group work exercise) during lectures where they are required to discuss and compare their findings of the AI-generated solutions to a case study or our case study assessments in Appendix A – C (if related). Furthermore, students can critique AI-generated solutions to research essay questions during lecture sessions. Moreover, the use of question banks can be helpful. In this scenario, students can generate questions on a particular topic to assess what they have learnt. Thus, these questions can be used for quick formative quizzes during the lecture session. To conclude, we recommend the use of different assessment tools ranging from in-person class test, which can be conducted online or through a written test, contextualised authentic assessments that are reflection-based, discussion-based assessment, and laboratory-based assessment (where applicable) to improve on assessing student learning outcomes in the GenAI era.

https://www.abdn.ac.uk/students/academic-life/common-grading-scale.php

Wang, H., Wu, W., Dou, Z., He, L., & Yang, L. (2023). Performance and exploration of ChatGPT in medical examination, records and education in Chinese: Pave the way for medical AI. *International Journal of Medical Informatics, 177*, 105173. https://doi.org/10.1016/j.ijmedinf.2023.105173

Wu, T., He, S., Liu, J., Sun, S., Liu, K., Han, Q.-L., & Tang, Y. (2023). A brief overview of ChatGPT: The history, status quo and potential future development. *IEEE/CAA Journal of Automatica Sinica, 10*(5), 1122–1136. https://doi.org/10.1109/JAS.2023.123618

Zarb, M., McDermott, R., Martin, K., Young, T., & McGowan, J. (2023). Evaluating a pass/fail grading model in first year undergraduate computing. *2023 IEEE Frontiers in Education Conference (FIE),* 1–9. http://dx.doi.org/10.1109/FIE58773.2023.10343276

# Appendices

## Appendix A

BMS services is a small-scale healthcare company. You have been asked by the team to explore machine learning techniques to analyse and evaluate heart failure model performance to make informed decisions on patient's survival. For this purpose, you will make use of programming and statistical concepts for analysis, visualisation, and machine learning algorithms.

Please download the health failure clinical dataset via this link below
https://archive.ics.uci.edu/dataset/519/heart+failure+clinical+records

Specifically, you are required to perform the following tasks below.
i). Exploratory Data Analysis
You are required to check if there's any missing data. List appropriate data pre-processing steps. Perform required descriptive statistical analysis. Give detailed explanation of all processes.
ii). Classification I
Split the dataset on training and testing sets. You are required to fit all machine learning algorithms namely, Naïve Bayes, Logistic Regression, Support Vector Machine, Random Forest classifier, K-Nearest Neighbour and Multi-Layer Perceptron Neural Networks. Evaluate your models using test dataset and provide the confusion matrix for all models. Report and compare performance of the models in terms of accuracy, precision, recall and F1-Score. Draw conclusions and provide recommendations.
iii). Classification II
Investigate class imbalance problem by producing the plot of the target variable class distribution. If there is presence of class imbalance problem, use at least 2 techniques to balance the class distribution (Algorithm or Sampling technique). This means you will have a balanced dataset. Using the balanced dataset, you are required to build classification models using machine learning algorithms namely, Naïve Bayes, Logistic Regression, Support Vector Machine, Random Forest classifier, K-Nearest Neighbour and Multi-Layer Perceptron Neural Networks. Evaluate your models using test dataset and provide the confusion matrix for all models. Report and compare performance of the models in terms of accuracy, precision, recall and F1-Score. Compare your result with the result of II above. Draw conclusions and provide recommendations. Please provide justification for chosen methods.
iv). Conclusion
Critically review solutions ii). and iii). above. Which model will you recommend?

## Appendix B

Task Summary:
You are a data scientist at Premium Technologies and recently tasked with developing a ML pipeline for one of your Biomedical clients. You are to perform a text classification of these scientific papers into thyroid, colon, or lung. To simplify the problem, it is sufficient to only consider the first 600 characters of each paper text. Afterwards, evaluate the model's performance. Please download the medical text dataset – cancer doc classification provided via the link below.
Kaggle: https://www.kaggle.com/datasets/falgunipatel19/biomedical-text-publication-classification
The dataset consists of the following:
labels: labels describing the different types of cancer; provided in column '0'
text: free text provides in column 'a' and contains medical papers addressing one of the 3 cancer areas
Requirements
Please perform the following tasks in a well-documented Jupyter notebook:
1. Load medical text dataset.
2. Perform appropriate data pre-processing steps.
3. Implement a text classification approach by using a transformer-model (e.g. BERT model) to convert the text into an embedding space.
4. Train and execute a classification model and evaluate the model's performance.
5. Depending on your results, describe further optimisation steps that you would consider in order to improve the performance of your model.

Additional Information
• Although this is a synthetic dataset, you are to operate under the assumption that this is patient confidential data which cannot be shared with externals
• The objective of this challenge is not to develop the best-performing algorithm, but rather to understand your motivation for a specific approach/solution and to understand your thinking about further improvements.
Report
Prepare a short report (not exceeding two pages including figures) describing your thoughts on why specific steps/methods were used. You are required to submit your report and the codes / Jupyter notebook.

## Appendix C

In project performance assessment, quality, cost, and time are the three apexes of the Iron Triangle. Yet, value, safety, and scope are also major criteria. You have been requested to represent your organisation at the 'performance brief' meeting for Qukzome community health center. You are required to create a One-Slide PowerPoint presentation that will illustrate schematically the interrelationship among the project criteria listed above using the Iron Triangle while emphasising the need to avoid or reduce dispute. In addition, give an appropriate title to the schematic illustration.

## Appendix D

| Assessment Criteria | Grade A (Excellent) | Grade B (Very Good) | Grade C (Good) | Grade D (Pass) | Grade E (Marginal Fail) | Grade F (Fail) | Grade G (Fail) |
|---|---|---|---|---|---|---|---|
| Comprehension and Application | Exceptional understanding with nuanced insights. Evidence of extensive reading / research. | Displays a very good understanding of the topic with comprehensive insights. Evidence of reading / research. | Good understanding with some depth and clarity. Evidence of research. | Basic understanding shown with general insights. Evidence of research. | Poor understanding of the topic. | Deficiencies in formulation of arguments which are sufficiently serious to indicate a fundamental lack of understanding of the assessment topic. | Token or no submission |
| Analysis and Synthesis, Accuracy and Relevance | Exceptional analysis of subject with use of advanced theory | Very good analysis of subject with use of relevant theory | Good analysis of subject with use of theory | Some analysis of subject with use of theory | Poor analysis of subject | No appreciable debate/ analysis | Token or no submission |
| Evaluation/Discussion of Result | Exceptional critical analysis with sophisticated integration of theory and examples. | Provides insightful analysis with well integrated, relevant theory and examples | Good level of analysis with relevant theory and examples. | Some analysis, but mostly descriptive or lacking depth. | Little to no critical analysis. | No appreciable evidence of coherent thinking | Token or no submission |
| Presentation/ Readability | Exceptional clarity, and elegance in writing, error-free | Exceptionally clear, concise, and error-free writing. | Clear writing with minor errors. | Writing is understandable but with several errors. | Poor writing quality with numerous errors. | Consistently poor writing quality/ presentation. | Token or no submission |